\documentclass[aps,prd,twocolumn,preprintnumbers,nofootinbib]{revtex4-2}
\usepackage[colorlinks=true, linkcolor=blue, citecolor=blue, urlcolor=blue]{hyperref}
\usepackage{bm,latexsym,amssymb,amsmath,mathrsfs}
\usepackage{graphicx}
\usepackage{color}
\usepackage{times,setspace}
\usepackage{physics}

\begin{document}

\title{$\mathbb{Z}_3$ lattice gauge theory as a toy model for dense QCD}

\author{Yoshimasa Hidaka}
\affiliation{Yukawa Institute for Theoretical Physics, Kyoto University, Kyoto 606-8502, Japan}
\affiliation{RIKEN iTHEMS, RIKEN, Wako 351-0198, Japan}

\author{Yuya Tanizaki}
\affiliation{Yukawa Institute for Theoretical Physics, Kyoto University, Kyoto 606-8502, Japan}

\author{Arata Yamamoto}
\affiliation{Department of Physics, The University of Tokyo, Tokyo 113-0033, Japan}

\begin{abstract}
    We propose the $(3+1)$-dimensional $\mathbb{Z}_3$ lattice gauge theory coupled with the 2-flavor Wilson-Dirac fermion as a toy model for studying quantum chromodynamics (QCD) at nonzero density. 
    We study its phase diagram in the space of the lattice gauge couplings $g^2$ and the quark chemical potentials $\mu$ and discuss the similarity and difference compared with anticipated behaviors of actual QCD. 
    This model also provides a testing ground for various algorithms of the numerical Hamiltonian formalism as its Hilbert space is finite-dimensional in a finite box. 
\end{abstract}

\preprint{YITP-24-13}

\maketitle

\section{Introduction}

Unveiling the dense state of nuclear matter remains a longstanding challenge in quantum chromodynamics (QCD)~\cite{Fukushima:2010bq}. 
Current theoretical calculations encounter limitations at nonzero density; for instance, nuclear many-body theory is restricted to low density, perturbative QCD to asymptotically high density, and lattice QCD simulations are hindered by the sign problem. 
Some anticipate that quantum computation may offer a future solution to this problem~\cite{Troyer:2004ge, Yamamoto:2021vxp, Tomiya:2022chr}.
The quantum computation of lattice QCD employs the Hamiltonian formalism, avoiding the sign problem associated with Monte Carlo sampling. 
However, it instead has to treat the exponentially huge Hilbert space. While the Hilbert space of fermion fields is finite-dimensional, that of gauge fields is infinite-dimensional, presenting a more formidable challenge. 
In practice, the local Hilbert space of gauge fields is truncated to finite dimensions, and the infinite-dimensional limit is subsequently taken. Nevertheless, the computational cost escalates rapidly in the infinite-dimensional limit \cite{Murairi:2022zdg}, posing one of the critical hurdles in quantum computing lattice QCD.

In this context, it is very desirable to have a toy model of nonzero-density QCD that has genuinely finite-dimensional Hilbert space, and 
we propose discrete gauge theories to accomplish the purpose. 
While lacking a well-defined continuum limit, discrete gauge theories prove valuable for testing quantum algorithms for field theories. 
The $\mathbb{Z}_2$ gauge theories are the most economical model but have only mesons, just like $SU(2)$ gauge theory.
Baryons are essential to imitate nonzero-density QCD, 
and thus the $\mathbb{Z}_3$ gauge theories will be the simplest candidate of more sensible models as they have both baryons and mesons. 
The mass spectrum of such a model has been studied in the one spatial dimension~\cite{Pisarski:2021aoz, Florio:2023kel}, and the model is termed as ``QZD'' (quantum $\mathbb{Z}_3$ dynamics) like QCD~\cite{Florio:2023kel}.
(For other uses of $\mathbb{Z}_3$ gauge theories, see, e.g., Refs.~\cite{Ercolessi:2017jbi, Magnifico:2019kyj} for the Schwinger model and Refs.~\cite{Gattringer:2012jt, Akiyama:2023hvt} for the Abelian-Higgs model.)
In this paper, we uncover the basic properties of the $(3+1)$-dimensional $\mathbb{Z}_3$ lattice gauge theory coupled to the $2$-flavor Wilson-Dirac fermion, and show that it has a rich phase diagram.

\section{Hamiltonian of 3+1d QZD}
\label{secB}

We consider a cubic lattice with periodic boundary conditions.
The gauge field lives on lattice links and the fermion is defined on lattice sites.
The gauge link operator $U_k (\bm{x})$ and its conjugate operator $\Pi_k (\bm{x})$ do not commute but obey
\begin{equation}
 \Pi_k (\bm{x}) U_k (\bm{x}) \Pi^\dagger_k (\bm{x}) = e^{\frac{2\pi i}{3}} U_k (\bm{x})
\end{equation}
with $k=1,2,3$ \cite{Horn:1979fy}.
For example, in the basis where the gauge link operator is diagonalized, these operators are written by the $3\times 3$ clock and shift matrices,
\begin{equation}
U_k (\bm{x}) =
    \begin{pmatrix}
   1 & 0 & 0 \\
   0 & e^{\frac{2\pi i}{3}} & 0 \\
   0 & 0 & e^{\frac{4\pi i}{3}} \\
    \end{pmatrix}
, \quad \Pi_k (\bm{x}) =
    \begin{pmatrix}
   0 & 1 & 0 \\
   0 & 0 & 1 \\
   1 & 0 & 0 \\
    \end{pmatrix}. 
\end{equation}
The fermion creation and annihilation operators obey the canonical anticommutation relation
\begin{equation}
 \left\{ \psi_\alpha({\bm x}) , \psi^\dagger_\beta({\bm y}) \right\} =  \delta_{\bm{x},\bm{y}}\delta_{\alpha,\beta},
\end{equation}
where $\alpha$ is a flavor-spinor index ($\alpha=1,\cdots,4N_f$).
The total Hilbert space including unphysical states is given by $(\mathbb{C}^3)^{3V}\otimes (\mathbb{C}^2)^{4N_f V}$ 
when the lattice volume is $V$, and its dimension is exponentially large, $3^{3V}\times 2^{4N_fV}$, but finite.
The lattice unit is used throughout the paper.

The Hamiltonian consists of electric, magnetic, and quark terms
\begin{equation}
 H = H_E + H_B + H_f.
\end{equation}
The electric and magnetic terms are
\begin{align}
    H_E &= \sum_{\bm{x}} \sum_k g^2 \left[1 - \frac{1}{2} \left\{ \Pi_k (\bm{x}) + \Pi^\dagger_k (\bm{x}) \right\}\right] ,\\
    H_B &= \sum_{\bm{x}} \sum_{k<l} \frac{1}{g^2} \left[1 - \frac{1}{2}  \left\{ U_{kl}(\bm{x}) + U_{kl}^\dagger(\bm{x}) \right\}\right] ,
\end{align}
where $U_{kl}(\bm{x})$ is the plaquette operator and $g^2$ is the lattice coupling constant, as usual.
We use $\Pi_k (\bm{x})$, instead of the electric field operator $E_k(\bm{x})$, for $H_E$ (though there is no essential difference in discrete gauge theories).
As for the quark part, we consider the $N_f=2$ Wilson-Dirac fermion
\begin{equation}
\begin{split}
H_f &= \sum_{\bm{x}} \Big[ (3r+m)  \psi^\dagger(\bm{x}) \gamma^0 \psi(\bm{x}) \\
&\quad -\frac{1}{2} \sum_k \Big\{ \psi^\dagger(\bm{x}) \gamma^0 (r-i \gamma^k) U_k(\bm{x}) \psi(\bm{x}+\bm{e}_k) \\
&\quad + \psi^\dagger(\bm{x}+\bm{e}_k) \gamma^0 (r+i \gamma^k) U^\dagger_k(\bm{x}) \psi(\bm{x}) \Big\}  \Big],
\end{split}
\end{equation}
where $m>0$ is the flavor-degenerate fermion mass and $r$ is the Wilson parameter to gap out the fermion doublers. 
The flavor-spinor index $\alpha$ is implicitly summed up. 
The quark number operator is defined by
\begin{equation}
 Q= \sum_{\bm{x}} \rho(\bm{x}) = \sum_{\bm{x}} \{ \psi^\dagger(\bm{x}) \psi(\bm{x}) -2N_f \}, 
\end{equation}
and it is conserved because of the commutation relation $[H,Q]=0$. 
At a nonzero chemical potential, the Hamiltonian is shifted by the conserved charge, $H -\mu Q$.
To circumvent the unphysical contributions from fermion doublers, the chemical potential needs to be much smaller than the Wilson parameter, $|\mu|\ll 2r$.\footnote{When we increase the chemical potential beyond $m+2r$, the new Fermi surface is formed due to the fermion doubler at a certain chemical potential. Moreover, the original Fermi surface and the Fermi surface of the doubler merge into the single Fermi surface with nontrivial topology for larger $\mu$. Therefore, this lattice model has many quantum phase transitions associated with the topology change of Fermi surface in the range $\mu\gtrsim 2r$.} 

The physical states, $|\Psi_\mathrm{phys}\rangle$, must satisfy the Gauss law at each site. 
The Gauss law constraint is written as
\begin{equation}
 \prod_{k=1,2,3} \Pi_k(\bm{x}) \Pi^\dagger_k(\bm{x}-\bm{e}_k) |\Psi_\mathrm{phys}\rangle= 
 e^{\frac{2\pi i}{3} \rho(\bm{x})}|\Psi_\mathrm{phys}\rangle . 
 \label{eq:Gauss}
\end{equation}
This equation means that the gauge charge is defined by $\rho(\bm{x})/3$ mod 1. 
If one would like to introduce a background static charge $q(\bm{x})$, one needs to replace $\rho(\bm{x})$ by $\rho(\bm{x})+q(\bm{x})$ on the right-hand-side of Eq.~\eqref{eq:Gauss}. 
By taking the products of Eq.~\eqref{eq:Gauss} for all sites $\bm{x}$, we find that $Q$ must be quantized in the multiples of $3$ for physical states, and we can define the baryon number by $B=Q/3$.

\section{Vacuum properties}
\label{sec:vacuum}

Let us first discuss the physics of $(3+1)$-dimensional QZD at the vacuum $\mu=0$ and the associated low-energy spectrum. 
When quarks decouple ($m\to \infty$), the system has a confined phase at the strong couplings $g^2\gg 1$ and a deconfined phase at the weak couplings $g^2 \ll 1$, and they are separated by quantum phase transitions. 
The deconfined phase is described by a topological order, which is stable under any local perturbations due to long-range entanglement.
The phase transition is first order in pure $\mathbb{Z}_3$ gauge theory \cite{Gattringer:2012jt}.
This suggests that a first-order phase transition at $g^2_c=O(1)$ is present even when finite-mass quarks are turned on, so we need to study the strong and weak couplings separately.

\subsection{Strong coupling}
\label{secS}

When $g^2\gg 1$ and $m\to \infty$, we can apply the strong-coupling expansion expansion to demonstrate the quark confinement analytically. 
In the strong coupling limit, the electric term in the Hamiltonian, $H_E$, is dominant.
Taking the basis diagonalizing $\Pi_k(\bm{x})$,
\begin{equation}
    \Pi_k(\bm{x}) |E_k(\bm{x})\rangle = e^{\frac{2\pi i}{3} E_k(\bm{x})}|E_k(\bm{x})\rangle 
\end{equation}
for the gauge part, 
the vacuum $|\Omega\rangle$ prefers the trivial configuration
\begin{equation}
    |\Omega\rangle = 
    \bigotimes_{k,\bm{x}}|E_{k}(\bm{x})=0\rangle. 
\end{equation}
The strong-coupling vacuum is thus unique, and its energy is $\varepsilon_0 = \langle\Omega| H_E |\Omega\rangle = 0$.
When there is a pair of static charge and anticharge, $q(\bm{x}_1)=1$ and $q(\bm{x}_2)=-1$ on the $x_k$ axis, the Gauss law forces the electric field to be nonzero along the minimal path connecting $\bm{x}_1$ and $\bm{x}_2$,
\begin{equation}
    E_k(\bm{x})=1 \quad \mathrm{for} \ \bm{x}=\bm{x}_1, \cdots, \bm{x}_2-\bm{e}_k.
\end{equation}
The system energy is proportional to the path length $\ell$ as $\varepsilon_0 = g^2(1- \cos\frac{2\pi}{3} )\ell$ \cite{Kogut:1979wt}.
We can show that any charge-neutral configuration of static charges, e.g., three charges, exhibits the same behavior.
This is nothing but the quark confinement in the strong-coupling limit.

\begin{figure}[t]
\begin{center}
\includegraphics[width=0.45\textwidth]{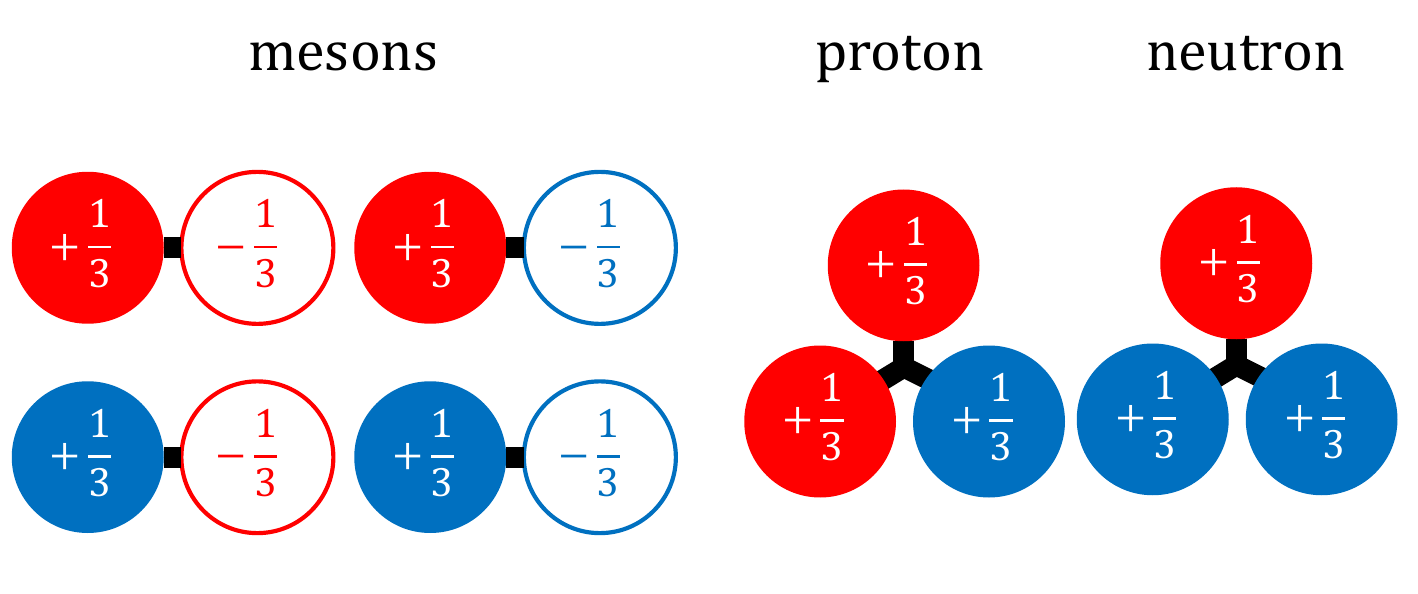}
\end{center}
\caption{Mesons and baryons in 2-flavor QZD.
The red and blue balls are $u$ and $d$ quarks, respectively.}
\label{fighadron}
\end{figure}

As a consequence, when we turn on quarks with a finite mass $m>0$, the low-energy spectrum is described by the tightly-bound charge-neutral objects (i.e., hadrons). 
A single quark has a nonzero charge but a meson and a baryon are charge neutral as drawn in Fig.~\ref{fighadron}.
As we consider the two-flavor case ($u$ and $d$), the ground states of these hadrons can exist on one lattice site and thus have finite masses in the strong coupling limit $g^2\to \infty$.
There are four degenerate ground-state baryons: protons ($uud$) and neutrons ($udd$) with spin $\pm 1/2$.
The low-energy hadron spectrum is quite similar to that of two-flavor QCD.

Explicit computations of hadron masses require higher-order calculations of the strong coupling expansion.
The second-order perturbation leads to spontaneous chiral symmetry breaking and dressed hadron masses $m+O(1/g^2)$~\cite{Smit:1980nf}.
When the quark mass $m$ is small, a pseudo-scalar meson is the pseudo-Nambu-Goldstone boson, so the lightest particle in the vacuum.\footnote{The same conclusion can be obtained by considering the path-integral formulation of this model. 
As the model is a vector-like gauge theory, its path integral at $\mu=0$ is semi-positive definite, and thus we can derive the analogue of QCD inequalities~\cite{Weingarten:1983uj, Vafa:1983tf, Witten:1983ut}. 
This immediately tells that the pseudo-scalar meson is the lightest in the confining vacuum.}
Therefore, we can conclude that the baryon-baryon interaction should be dominated by the pseudo-scalar-meson exchange at long distances and thus it is attractive.

\subsection{Weak coupling}
\label{secW}

Next, we consider the weak coupling region, $g^2\ll 1$, with $m\to \infty$. 
In this limit, the Hamiltonian becomes $H = H_B$.
Let us first find the ground state.\footnote{One of the ground states at the weak coupling limit can be constructed by using  the wavefunction of the ground state at the strong coupling limit as
\begin{equation}
    |\Omega_g\rangle=
    \prod_{\bm{x}}\prod_{k<l} U^{(r)}_{kl} (\bm{x})\bigotimes_{\bm{x},k}|E_k(\bm{x})=0\rangle
    \nonumber
\end{equation}
where we introduce
$
    U^{(r)}_{kl} (\bm{x}) = \frac{1}{3}(1+ U_{kl}(\bm{x})+ U_{kl}^2(\bm{x})),
    \nonumber
$
which satisfies $U_{kl} (\bm{x})U^{(r)}_{kl} (\bm{x})=U^{(r)}_{kl} (\bm{x})$.
When the spatial manifolds have a nontrivial cycle, the above one corresponds to the symmetric sum of degenerate ground-state wavefunctions. 
} The Hamiltonian is minimized if 
\begin{equation}
     U_{kl}(\bm{x}) |\Omega\rangle = |\Omega\rangle
     \label{eq:Omega_g}
\end{equation}
for all plaquettes $U_{kl}(\bm{x})$. 
This condition does not specify the value of Wilson loops wrapping nontrivial cycles of the spatial manifold, and their possible values determine the ground-state degeneracy. 
For example, the ground state is unique on the $3$-sphere $S^3$, while the ground states are $3^{3}=27$-fold degenerate on the $3$-torus $T^3$. 
This is the manifestation of long-range entanglement, and the low-energy effective theory becomes the topological field theory~\cite{Wen:1989zg, Kitaev:2005hzj}. 

When the quark mass $m$ is finite, there is a contribution of the fermionic excitation that travels around a nontrivial spatial cycle, so the $27$-fold degeneracy in the above discussion is lifted by $O(e^{-mL})$ at finite volume, where $L$ is the size of the torus. 
This shows that the topology dependence of the ground-state degeneracy still exists in the infinite-volume limit, $L\to \infty$. 
In general, the topological order is stable under any local perturbations, and thus the weak-coupling regime is still separated from the strong-coupling regime by some quantum phase transitions. 

In this regime, the system is ``deconfined''. 
Of course, the Gauss law is violated if we apply the single-quark operator to the vacuum, but we can consider the separated quark and anti-quark connected by the Wilson line, instead, as a physical state. 
The energy does not cost at all even if they are arbitrarily separated, and we may regard, in this sense, that low-energy excitations are well described by free quarks and antiquarks.

\section{Phase diagram}
\label{secP}

So far, our discussion has focused on the properties of the vacuum. 
We would like to extend our discussion to a nonzero-density system by introducing the quark chemical potential $\mu$, whose ground state minimizes $H -\mu Q$.

\begin{figure}[t]
\begin{center}
\includegraphics[width=0.5\textwidth]{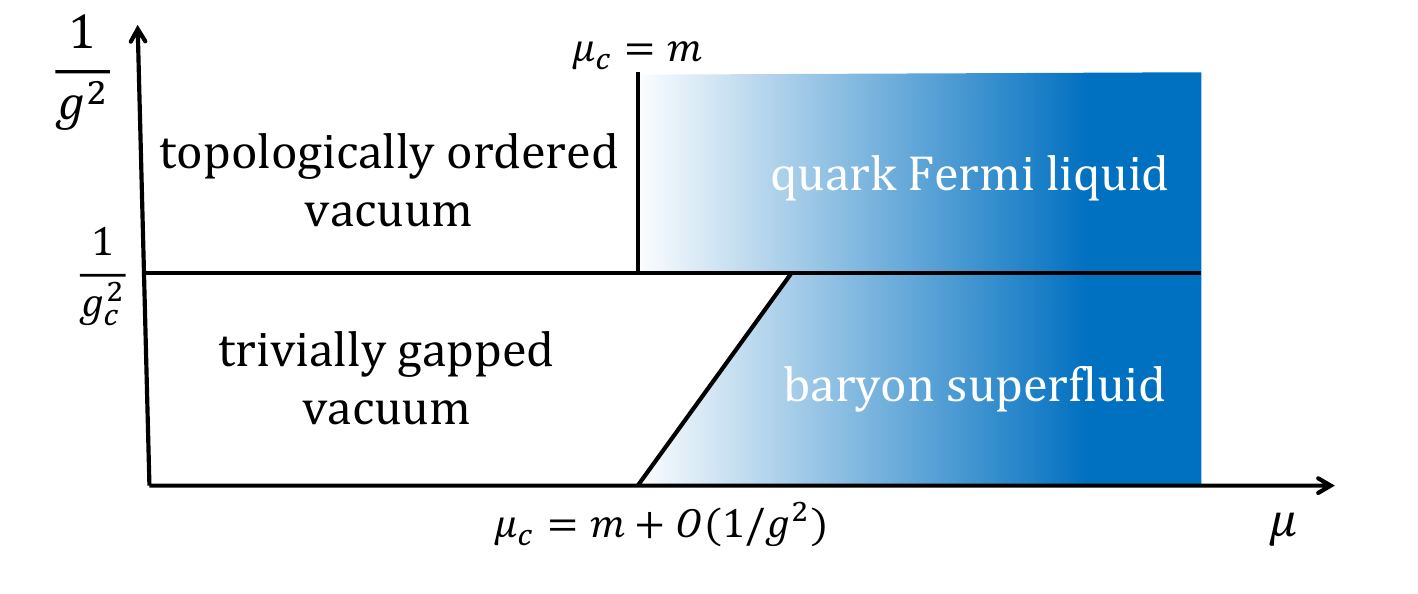}
\end{center}
\caption{Conjectured phase diagram of (3+1)-dimensional QZD. In the vacuum, the strong- and weak-coupling regimes are separated by the confinement-deconfinement phase transition. At nonzero densities, quarks form a stable Fermi surface in the weak coupling, while baryons form a superfluid condensate in the strong coupling. So there are at least four distinct phases. }
\label{figPD}
\end{figure}

When $\mu$ is small enough, quark/baryon excitations are suppressed and the ground-state wave function remains the vacuum one. 
In the context of the sign problem for nonzero-density QCD, this is sometimes called the Silver Blaze phenomenon~\cite{Cohen:2003kd, Nagata:2012tc, Tanizaki:2015rda}. 
As we have seen, the system is gapped and the strong-coupling and weak-coupling regimes are separated by the quantum phase transition. 

Let us increase the chemical potential and consider the situation where fermions start to be occupied in the vacuum. In the weak-coupling case, if $\mu>\mu_c= m$, the quark Fermi surface is formed and the low-energy physics is well-described by the Fermi liquid theory. 
The interaction between quarks is given by exchanging the massive Abelian gauge field, and thus the particle-particle interaction is repulsive. 
Therefore, there is no Cooper instability and the Fermi surface of quarks is stable, so the system has gapless fermionic excitations.  
The phase transition from the vacuum to the quark Fermi liquid will be second-order.

In the strong coupling case, low-energy excitations are gauge-singlet hadrons, and only baryons are charged under the $U(1)$ symmetry. 
Therefore, the chemical potential must exceed the $1/3$ of the baryon mass $M$, i.e., $\mu>\mu_c=M/3\sim m+O(1/g^2)$, in order to have the Fermi surface of baryons.
The strong coupling expansion predicts that the onset of the baryon matter is a first-order phase transition~\cite{Umino:2001ie,Fang:2002rk}.
We have seen that the baryon-baryon interaction is attractive at long distances in the vacuum case, and it would be natural to expect that this attraction survives even at nonzero densities. 
If so, the Fermi surface of baryons has the Cooper instability. 
As a result, baryon number symmetry is spontaneously broken and baryon superfluidity is formed. 

Summarizing these results, there are at least four distinct phases; confinement (trivially gapped) vacuum, deconfinement (topologically ordered) vacuum, quark Fermi liquid, and baryon superfluid.  
The conjectured phase diagram is given in Fig.~\ref{figPD}, but it is still sketchy and
numerical effort is necessary to determine the locations and the orders of phase transitions. 
Note that we have not considered phase transition lines associated with the spontaneous breaking of chiral symmetry, as we are focusing on quarks with finite mass. 
For small quark masses, phase transition lines may arise due to approximate chiral symmetry.

\section{Discussion}
\label{sec:discussion}

As we have determined the phase structure of QZD, we would like to discuss what kind of lessons can be learned from this model for nonzero-density QCD. 
In particular, we clarify the similarities and differences between our model and actual QCD, and point the potential direction to make their dynamics more similar. 

The most significant similarity is the existence of baryons as charge-neutral composite particles.
In the strong coupling, quarks are confined into hadrons, and their low-energy feature is similar to the one in actual QCD. 
We may think that $\mathbb{Z}_3$ corresponds to the center of the $SU(3)$ gauge group, and then the confinement of this model is analogous to the center-vortex scenario for the confinement mechanism~\cite{Cornwall:1979hz, Ambjorn:1980ms, Greensite:2003bk, Tanizaki:2022ngt}. 
This may be the reason why the properties of the hadron phase survive to some extent for discrete gauge theories. 

Lastly, let us point out a few important differences when QZD is compared with actual QCD:
\begin{description}
    \item[Absence of the asymptotic freedom] 
In actual QCD, the gauge coupling depends on the energy scale and it becomes weaker at high densities.
This feature cannot be realized in discrete gauge theories, and thus the hadron phase does not go to the quark phase even if we increase the chemical potential. 
Therefore, we need to change the gauge coupling directly to discuss the connection between the hadron and quark phases. 
This difference cannot be circumvented as long as we use discrete gauge theories. 
    \item[Absence of the continuum limit]
Lattice theories without the asymptotic freedom do not have the well-defined continuum limit.
Such theories cannot reproduce the quantum anomaly that comes from the ultraviolet divergence in continuous spacetime.
The model must be modified to mimic the anomaly for studying anomaly-induced phenomena, such as the large mass of the eta meson.
    \item[Quark-quark interaction]
In $SU(3)$, the one-gluon exchange between quarks has two channels, $\bm{3}\otimes \bm{3}=\bm{6}\oplus \bar{\bm{3}}$. 
The symmetric channel $\bm{6}$ is repulsive, but the antisymmetric channel $\bar{\bm{3}}$ is attractive. 
Due to this attraction, the quark Fermi sea gets the Cooper instability and the diquark condensate is formed in the mean-field analysis, which causes the color superconductivity~\cite{Alford:1997zt}. 
In our model, however, the quark Fermi surface is stable because the quark-quark interaction is repulsive for Abelian gauge theories. 
To cure this issue, we need to have the counterpart of the $\bar{\bm{3}}$ channel by extending the gauge group from $\mathbb{Z}_3$ to some non-Abelian finite groups, or even more exotic objects such as quantum groups~\cite{Zache:2023dko, Hayata:2023puo, Hayata:2023bgh}. 
Those extensions of our model may provide an explicit realization for the quark-hadron continuity in the $3$-flavor situation~\cite{Schafer:1998ef, Hatsuda:2006ps, Yamamoto:2007ah}. 
The validity of quark-hadron continuity in nonzero-density QCD is recently debated due to the nontrivial feature of the Aharonov-Bohm phase around the superfluid vortex~\cite{Cherman:2018jir, Hirono:2018fjr, Hirono:2019oup, Cherman:2020hbe, Hidaka:2022blq, Hayashi:2023sas, Cherman:2024exo}, so it would be nice to develop various models for its concrete establishment in QCD-like theories.
\end{description}

\begin{acknowledgments}
This work was started in the discussion during the YITP long-term workshop, ``Quantum Information, Quantum Matter and Quantum Gravity'' (YITP-T-23-01), and the authors thank the organizers for providing the nice opportunity. 
This work was supported by JSPS KAKENHI Grant No.~21H01084, 24H00975 (Y.~H.), 22H01218, 23K22489 (Y.~T.), and 19K03841 (A.~Y.).
The work of Y.~H. and Y.~T. was also partially supported by Center for Gravitational Physics and Quantum Information (CGPQI) at Yukawa Institute for Theoretical Physics.
\end{acknowledgments}

\bibliographystyle{utphys}
\bibliography{paper}

\providecommand{\href}[2]{#2}\begingroup\raggedright\begin{thebibliography}{10}

\bibitem{Fukushima:2010bq}
K.~Fukushima and T.~Hatsuda, ``{The phase diagram of dense QCD},''
  \href{http://dx.doi.org/10.1088/0034-4885/74/1/014001}{{\em Rept. Prog.
  Phys.} {\bfseries 74} (2011) 014001},
  \href{http://arxiv.org/abs/1005.4814}{{\ttfamily arXiv:1005.4814 [hep-ph]}}.

\bibitem{Troyer:2004ge}
M.~Troyer and U.-J. Wiese, ``{Computational complexity and fundamental
  limitations to fermionic quantum Monte Carlo simulations},''
  \href{http://dx.doi.org/10.1103/PhysRevLett.94.170201}{{\em Phys. Rev. Lett.}
  {\bfseries 94} (2005) 170201},
  \href{http://arxiv.org/abs/cond-mat/0408370}{{\ttfamily
  arXiv:cond-mat/0408370}}.

\bibitem{Yamamoto:2021vxp}
A.~Yamamoto, ``{Quantum variational approach to lattice gauge theory at nonzero
  density},'' \href{http://dx.doi.org/10.1103/PhysRevD.104.014506}{{\em Phys.
  Rev. D} {\bfseries 104} (2021) 014506},
  \href{http://arxiv.org/abs/2104.10669}{{\ttfamily arXiv:2104.10669
  [hep-lat]}}.

\bibitem{Tomiya:2022chr}
A.~Tomiya, ``{Schwinger model at finite temperature and density with beta
  VQE},'' \href{http://arxiv.org/abs/2205.08860}{{\ttfamily arXiv:2205.08860
  [hep-lat]}}.

\bibitem{Murairi:2022zdg}
E.~M. Murairi, M.~J. Cervia, H.~Kumar, P.~F. Bedaque, and A.~Alexandru, ``{How
  many quantum gates do gauge theories require?},''
  \href{http://dx.doi.org/10.1103/PhysRevD.106.094504}{{\em Phys. Rev. D}
  {\bfseries 106} no.~9, (2022) 094504},
  \href{http://arxiv.org/abs/2208.11789}{{\ttfamily arXiv:2208.11789
  [hep-lat]}}.

\bibitem{Pisarski:2021aoz}
R.~D. Pisarski, ``{Remarks on nuclear matter: How an $\omega_0$ condensate can
  spike the speed of sound, and a model of $Z(3)$ baryons},''
  \href{http://dx.doi.org/10.1103/PhysRevD.103.L071504}{{\em Phys. Rev. D}
  {\bfseries 103} no.~7, (2021) L071504},
  \href{http://arxiv.org/abs/2101.05813}{{\ttfamily arXiv:2101.05813
  [nucl-th]}}.

\bibitem{Florio:2023kel}
A.~Florio, A.~Weichselbaum, S.~Valgushev, and R.~D. Pisarski, ``{Mass gaps of a
  $\mathbb{Z}_3$ gauge theory with three fermion flavors in 1 + 1
  dimensions},'' \href{http://arxiv.org/abs/2310.18312}{{\ttfamily
  arXiv:2310.18312 [hep-th]}}.

\bibitem{Ercolessi:2017jbi}
E.~Ercolessi, P.~Facchi, G.~Magnifico, S.~Pascazio, and F.~V. Pepe, ``{Phase
  Transitions in $Z_{n}$ Gauge Models: Towards Quantum Simulations of the
  Schwinger-Weyl QED},''
  \href{http://dx.doi.org/10.1103/PhysRevD.98.074503}{{\em Phys. Rev. D}
  {\bfseries 98} no.~7, (2018) 074503},
  \href{http://arxiv.org/abs/1705.11047}{{\ttfamily arXiv:1705.11047
  [quant-ph]}}.

\bibitem{Magnifico:2019kyj}
G.~Magnifico, M.~Dalmonte, P.~Facchi, S.~Pascazio, F.~V. Pepe, and
  E.~Ercolessi, ``{Real Time Dynamics and Confinement in the $\mathbb{Z}_{n}$
  Schwinger-Weyl lattice model for 1+1 QED},''
  \href{http://dx.doi.org/10.22331/q-2020-06-15-281}{{\em Quantum} {\bfseries
  4} (2020) 281}, \href{http://arxiv.org/abs/1909.04821}{{\ttfamily
  arXiv:1909.04821 [quant-ph]}}.

\bibitem{Gattringer:2012jt}
C.~Gattringer and A.~Schmidt, ``{Gauge and matter fields as surfaces and loops
  - an exploratory lattice study of the Z(3) Gauge-Higgs model},''
  \href{http://dx.doi.org/10.1103/PhysRevD.86.094506}{{\em Phys. Rev. D}
  {\bfseries 86} (2012) 094506},
  \href{http://arxiv.org/abs/1208.6472}{{\ttfamily arXiv:1208.6472 [hep-lat]}}.

\bibitem{Akiyama:2023hvt}
S.~Akiyama and Y.~Kuramashi, ``{Critical endpoint of (3+1)-dimensional finite
  density \ensuremath{\mathbb{Z}}$_{3}$ gauge-Higgs model with tensor
  renormalization group},''
  \href{http://dx.doi.org/10.1007/JHEP10(2023)077}{{\em JHEP} {\bfseries 10}
  (2023) 077}, \href{http://arxiv.org/abs/2304.07934}{{\ttfamily
  arXiv:2304.07934 [hep-lat]}}.

\bibitem{Horn:1979fy}
D.~Horn, M.~Weinstein, and S.~Yankielowicz, ``{Hamiltonian approach to Z(N)
  lattice gauge theories},''
  \href{http://dx.doi.org/10.1103/PhysRevD.19.3715}{{\em Phys. Rev. D}
  {\bfseries 19} (1979) 3715}.

\bibitem{Kogut:1979wt}
J.~B. Kogut, ``{An Introduction to Lattice Gauge Theory and Spin Systems},''
  \href{http://dx.doi.org/10.1103/RevModPhys.51.659}{{\em Rev. Mod. Phys.}
  {\bfseries 51} (1979) 659}.

\bibitem{Smit:1980nf}
J.~Smit, ``{Chiral Symmetry Breaking in QCD: Mesons as Spin Waves},''
  \href{http://dx.doi.org/10.1016/0550-3213(80)90056-5}{{\em Nucl. Phys. B}
  {\bfseries 175} (1980) 307--348}.

\bibitem{Weingarten:1983uj}
D.~Weingarten, ``{Mass Inequalities for QCD},''
  \href{http://dx.doi.org/10.1103/PhysRevLett.51.1830}{{\em Phys. Rev. Lett.}
  {\bfseries 51} (1983) 1830}.

\bibitem{Vafa:1983tf}
C.~Vafa and E.~Witten, ``{Restrictions on Symmetry Breaking in Vector-Like
  Gauge Theories},'' \href{http://dx.doi.org/10.1016/0550-3213(84)90230-X}{{\em
  Nucl. Phys. B} {\bfseries 234} (1984) 173--188}.

\bibitem{Witten:1983ut}
E.~Witten, ``{Some Inequalities Among Hadron Masses},''
  \href{http://dx.doi.org/10.1103/PhysRevLett.51.2351}{{\em Phys. Rev. Lett.}
  {\bfseries 51} (1983) 2351}.

\bibitem{Wen:1989zg}
X.~G. Wen, ``{Vacuum Degeneracy of Chiral Spin States in Compactified Space},''
  \href{http://dx.doi.org/10.1103/PhysRevB.40.7387}{{\em Phys. Rev. B}
  {\bfseries 40} (1989) 7387--7390}.

\bibitem{Kitaev:2005hzj}
A.~Kitaev, ``{Anyons in an exactly solved model and beyond},''
  \href{http://dx.doi.org/10.1016/j.aop.2005.10.005}{{\em Annals Phys.}
  {\bfseries 321} no.~1, (2006) 2--111},
  \href{http://arxiv.org/abs/cond-mat/0506438}{{\ttfamily
  arXiv:cond-mat/0506438}}.

\bibitem{Cohen:2003kd}
T.~D.~. Cohen, ``{Functional integrals for QCD at nonzero chemical potential
  and zero density},''
  \href{http://dx.doi.org/10.1103/PhysRevLett.91.222001}{{\em Phys. Rev. Lett.}
  {\bfseries 91} (2003) 222001},
  \href{http://arxiv.org/abs/hep-ph/0307089}{{\ttfamily arXiv:hep-ph/0307089}}.

\bibitem{Nagata:2012tc}
{\bfseries XQCD-J} Collaboration, K.~Nagata, S.~Motoki, Y.~Nakagawa,
  A.~Nakamura, and T.~Saito, ``{Towards extremely dense matter on the
  lattice},'' \href{http://dx.doi.org/10.1093/ptep/pts003}{{\em PTEP}
  {\bfseries 2012} (2012) 01A103},
  \href{http://arxiv.org/abs/1204.1412}{{\ttfamily arXiv:1204.1412 [hep-lat]}}.

\bibitem{Tanizaki:2015rda}
Y.~Tanizaki, Y.~Hidaka, and T.~Hayata, ``{Lefschetz-thimble analysis of the
  sign problem in one-site fermion model},''
  \href{http://dx.doi.org/10.1088/1367-2630/18/3/033002}{{\em New J. Phys.}
  {\bfseries 18} no.~3, (2016) 033002},
  \href{http://arxiv.org/abs/1509.07146}{{\ttfamily arXiv:1509.07146
  [hep-th]}}.

\bibitem{Umino:2001ie}
Y.~Umino, ``{Hamiltonian lattice QCD at finite density: Equation of state in
  the strong coupling limit},''
  \href{http://dx.doi.org/10.1103/PhysRevD.66.074501}{{\em Phys. Rev. D}
  {\bfseries 66} (2002) 074501},
  \href{http://arxiv.org/abs/hep-ph/0101144}{{\ttfamily arXiv:hep-ph/0101144}}.

\bibitem{Fang:2002rk}
Y.-Z. Fang and X.-Q. Luo, ``{Hamiltonian lattice quantum chromodynamics at
  finite density with Wilson fermions},''
  \href{http://dx.doi.org/10.1103/PhysRevD.69.114501}{{\em Phys. Rev. D}
  {\bfseries 69} (2004) 114501},
  \href{http://arxiv.org/abs/hep-lat/0210031}{{\ttfamily
  arXiv:hep-lat/0210031}}.

\bibitem{Cornwall:1979hz}
J.~M. Cornwall, ``{Quark Confinement and Vortices in Massive Gauge Invariant
  QCD},'' \href{http://dx.doi.org/10.1016/0550-3213(79)90111-1}{{\em Nucl.
  Phys. B} {\bfseries 157} (1979) 392--412}.

\bibitem{Ambjorn:1980ms}
J.~Ambjorn and P.~Olesen, ``{A Color Magnetic Vortex Condensate in QCD},''
  \href{http://dx.doi.org/10.1016/0550-3213(80)90150-9}{{\em Nucl. Phys. B}
  {\bfseries 170} (1980) 265--282}.

\bibitem{Greensite:2003bk}
J.~Greensite, ``{The Confinement problem in lattice gauge theory},''
  \href{http://dx.doi.org/10.1016/S0146-6410(03)90012-3}{{\em Prog. Part. Nucl.
  Phys.} {\bfseries 51} (2003) 1},
  \href{http://arxiv.org/abs/hep-lat/0301023}{{\ttfamily
  arXiv:hep-lat/0301023}}.

\bibitem{Tanizaki:2022ngt}
Y.~Tanizaki and M.~\"Unsal, ``{Center vortex and confinement in
  Yang\textendash{}Mills theory and QCD with anomaly-preserving
  compactifications},'' \href{http://dx.doi.org/10.1093/ptep/ptac042}{{\em
  PTEP} {\bfseries 2022} no.~4, (2022) 04A108},
  \href{http://arxiv.org/abs/2201.06166}{{\ttfamily arXiv:2201.06166
  [hep-th]}}.

\bibitem{Alford:1997zt}
M.~G. Alford, K.~Rajagopal, and F.~Wilczek, ``{QCD at finite baryon density:
  Nucleon droplets and color superconductivity},''
  \href{http://dx.doi.org/10.1016/S0370-2693(98)00051-3}{{\em Phys. Lett. B}
  {\bfseries 422} (1998) 247--256},
  \href{http://arxiv.org/abs/hep-ph/9711395}{{\ttfamily arXiv:hep-ph/9711395}}.

\bibitem{Zache:2023dko}
T.~V. Zache, D.~Gonz\'alez-Cuadra, and P.~Zoller, ``{Quantum and Classical
  Spin-Network Algorithms for q-Deformed Kogut-Susskind Gauge Theories},''
  \href{http://dx.doi.org/10.1103/PhysRevLett.131.171902}{{\em Phys. Rev.
  Lett.} {\bfseries 131} no.~17, (2023) 171902},
  \href{http://arxiv.org/abs/2304.02527}{{\ttfamily arXiv:2304.02527
  [quant-ph]}}.

\bibitem{Hayata:2023puo}
T.~Hayata and Y.~Hidaka, ``{String-net formulation of Hamiltonian lattice
  Yang-Mills theories and quantum many-body scars in a nonabelian gauge
  theory},'' \href{http://dx.doi.org/10.1007/JHEP09(2023)126}{{\em JHEP}
  {\bfseries 09} (2023) 126}, \href{http://arxiv.org/abs/2305.05950}{{\ttfamily
  arXiv:2305.05950 [hep-lat]}}.

\bibitem{Hayata:2023bgh}
T.~Hayata and Y.~Hidaka, ``{q deformed formulation of Hamiltonian SU(3)
  Yang-Mills theory},'' \href{http://dx.doi.org/10.1007/JHEP09(2023)123}{{\em
  JHEP} {\bfseries 09} (2023) 123},
  \href{http://arxiv.org/abs/2306.12324}{{\ttfamily arXiv:2306.12324
  [hep-lat]}}.

\bibitem{Schafer:1998ef}
T.~Sch\"afer and F.~Wilczek, ``{Continuity of quark and hadron matter},''
  \href{http://dx.doi.org/10.1103/PhysRevLett.82.3956}{{\em Phys. Rev. Lett.}
  {\bfseries 82} (1999) 3956--3959},
  \href{http://arxiv.org/abs/hep-ph/9811473}{{\ttfamily arXiv:hep-ph/9811473}}.

\bibitem{Hatsuda:2006ps}
T.~Hatsuda, M.~Tachibana, N.~Yamamoto, and G.~Baym, ``{New critical point
  induced by the axial anomaly in dense QCD},''
  \href{http://dx.doi.org/10.1103/PhysRevLett.97.122001}{{\em Phys. Rev. Lett.}
  {\bfseries 97} (2006) 122001},
  \href{http://arxiv.org/abs/hep-ph/0605018}{{\ttfamily arXiv:hep-ph/0605018}}.

\bibitem{Yamamoto:2007ah}
N.~Yamamoto, M.~Tachibana, T.~Hatsuda, and G.~Baym, ``{Phase structure,
  collective modes, and the axial anomaly in dense QCD},''
  \href{http://dx.doi.org/10.1103/PhysRevD.76.074001}{{\em Phys. Rev. D}
  {\bfseries 76} (2007) 074001},
  \href{http://arxiv.org/abs/0704.2654}{{\ttfamily arXiv:0704.2654 [hep-ph]}}.

\bibitem{Cherman:2018jir}
A.~Cherman, S.~Sen, and L.~G. Yaffe, ``{Anyonic particle-vortex statistics and
  the nature of dense quark matter},''
  \href{http://dx.doi.org/10.1103/PhysRevD.100.034015}{{\em Phys. Rev. D}
  {\bfseries 100} no.~3, (2019) 034015},
  \href{http://arxiv.org/abs/1808.04827}{{\ttfamily arXiv:1808.04827
  [hep-th]}}.

\bibitem{Hirono:2018fjr}
Y.~Hirono and Y.~Tanizaki, ``{Quark-Hadron Continuity beyond the
  Ginzburg-Landau Paradigm},''
  \href{http://dx.doi.org/10.1103/PhysRevLett.122.212001}{{\em Phys. Rev.
  Lett.} {\bfseries 122} no.~21, (2019) 212001},
  \href{http://arxiv.org/abs/1811.10608}{{\ttfamily arXiv:1811.10608
  [hep-th]}}.

\bibitem{Hirono:2019oup}
Y.~Hirono and Y.~Tanizaki, ``{Effective gauge theories of superfluidity with
  topological order},'' \href{http://dx.doi.org/10.1007/JHEP07(2019)062}{{\em
  JHEP} {\bfseries 07} (2019) 062},
  \href{http://arxiv.org/abs/1904.08570}{{\ttfamily arXiv:1904.08570
  [hep-th]}}.

\bibitem{Cherman:2020hbe}
A.~Cherman, T.~Jacobson, S.~Sen, and L.~G. Yaffe, ``{Higgs-confinement phase
  transitions with fundamental representation matter},''
  \href{http://dx.doi.org/10.1103/PhysRevD.102.105021}{{\em Phys. Rev. D}
  {\bfseries 102} no.~10, (2020) 105021},
  \href{http://arxiv.org/abs/2007.08539}{{\ttfamily arXiv:2007.08539
  [hep-th]}}.

\bibitem{Hidaka:2022blq}
Y.~Hidaka and D.~Kondo, ``{Emergent higher-form symmetry in Higgs phases with
  superfluidity},'' \href{http://arxiv.org/abs/2210.11492}{{\ttfamily
  arXiv:2210.11492 [hep-th]}}.

\bibitem{Hayashi:2023sas}
Y.~Hayashi, ``{Higgs-confinement continuity and matching of Aharonov-Bohm
  phases},'' \href{http://arxiv.org/abs/2303.02129}{{\ttfamily arXiv:2303.02129
  [hep-th]}}.

\end{thebibliography}\endgroup

\end{document}